\documentclass[longauth]{aa}  

\usepackage{graphicx}
\usepackage{txfonts}
\usepackage{booktabs}
\usepackage{url}
\usepackage{xcolor}

\newcommand{\emth}[1]{\ensuremath{#1}\xspace}

\newcommand{\tstar}{TOI~263\xspace}
\newcommand{\tplanet}{TOI~263.01\xspace}
\newcommand{\tic}{TIC~120916706\xspace}
\newcommand{\pytransit}{\textsc{PyTransit}\xspace}
\newcommand{\ldtk}{\textsc{LDTk}\xspace}

\newcommand{\llh}{\ensuremath{\ln P}}

\newcommand{\pvec}{\ensuremath{\boldsymbol{\theta}}\xspace}
\newcommand{\covmat}{\emth{\boldsymbol{\Sigma}}}

\newcommand{\gcm}{\emth{\mathrm{g\,cm^{-3}}}}
\newcommand{\smass}{\emth{\mathrm{M_\star}}}
\newcommand{\srad}{\emth{\mathrm{R_\star}}}

\newcommand{\rptrue}{\ensuremath{r_\mathrm{p,true}}\xspace}
\newcommand{\ktrue}{\ensuremath{k_\mathrm{true}}\xspace}
\newcommand{\kapp}{\ensuremath{k_\mathrm{app}}\xspace}
\newcommand{\teff}{\ensuremath{T_\mathrm{Eff}}\xspace}
\newcommand{\teffh}{\ensuremath{T_\mathrm{Eff,H}}\xspace}
\newcommand{\teffc}{\ensuremath{T_\mathrm{Eff,C}}\xspace}

\newcommand{\pimp}{\ensuremath{b}\xspace}

\newcommand{\pcref}{\ensuremath{c}\xspace}

\newcommand{\mjup}{\ensuremath{M_\mathrm{Jup}}\xspace}
\newcommand{\rjup}{\ensuremath{R_\mathrm{Jup}}\xspace}

\newcommand{\msun}{\ensuremath{M_\odot}\xspace}
\newcommand{\rsun}{\ensuremath{R_\odot}\xspace}

\newcommand{\tess}{\textit{TESS}\xspace}

\newcommand{\gaia}{\textit{Gaia}\xspace}

\newcommand{\tmodel}{\ensuremath{\mathcal{T}}\xspace}
\newcommand{\UP}[1]{\ensuremath{\mathcal{U}(#1)}\xspace}
\newcommand{\NP}[1]{\ensuremath{\mathcal{N}(#1)}\xspace}

\newcommand{\ktmedian}{0.217\xspace}

\newcommand{\ktupper}{0.286\xspace}
\newcommand{\rtmedian}{0.87\xspace}
\newcommand{\rtlower}{0.44\xspace}
\newcommand{\rtupper}{1.41\xspace}

\newcommand{\giturl}{\url{https://github.com/hpparvi/parviainen_2019b_toi_263}}

\begin{document} 

   \title{MuSCAT2 multicolour validation of TESS candidates: an ultra-short-period substellar object around an M dwarf.}
   \titlerunning{MuSCAT2 validation of \tplanet}

   \author{H. Parviainen\inst{\ref{iiac},\ref{iull}}
   \and E. Palle\inst{\ref{iiac},\ref{iull}} 
   \and M.R. Zapatero-Osorio\inst{\ref{csic}} 
   \and P. Montanes Rodriguez\inst{\ref{iiac},\ref{iull}} 
   \and F. Murgas\inst{\ref{iiac},\ref{iull}} 
   \and N. Narita\inst{\ref{iutda},\ref{iabc},\ref{ijsta},\ref{inao},\ref{iuteps}} 
   \and D.~Hidalgo~Soto\inst{\ref{iiac},\ref{iull}}
   \and V. J. S.~B\'ejar\inst{\ref{iiac},\ref{iull}}
   \and J.~Korth\inst{\ref{ikoln}}
   \and M.~Monelli\inst{\ref{iiac},\ref{iull}} 
   \and N.~Casasayas~Barris\inst{\ref{iiac},\ref{iull}}
   \and N.~Crouzet\inst{\ref{iesa}}
   \and J.P.~de Leon\inst{\ref{iutda}}
   \and A.~Fukui\inst{\ref{iuteps}}
   \and A.~Hernandez\inst{\ref{iiac},\ref{iull}}
   \and P.~Klagyivik\inst{\ref{iiac},\ref{iull}}
   \and N.~Kusakabe \inst{\ref{iabc},\ref{inao}} 
   \and R.~Luque\inst{\ref{iiac},\ref{iull}}
   \and M.~Mori\inst{\ref{iutda}}
   \and T.~Nishiumi\inst{\ref{ikyo}}
   \and J.~Prieto-Arranz\inst{\ref{iiac},\ref{iull}}
   \and M.~Tamura\inst{\ref{iutda},\ref{iabc},\ref{inao}}
   \and N.~Watanabe\inst{\ref{inao}}
   \and C.~Burke\inst{\ref{imit}}
   \and D.~Charbonneau\inst{\ref{icfa}}
   \and K.A.~Collins\inst{\ref{icfa}}
   \and K.I.~Collins\inst{\ref{ivandy}}
   \and D. Conti\inst{\ref{iaavso}}
   \and A.~Garcia~Soto\inst{\ref{iags}}
   \and J.S.~Jenkins\inst{\ref{iuc}}
   \and J.M.~Jenkins\inst{\ref{iames}}
   \and A.~Levine\inst{\ref{imit}}
   \and J.~Li\inst{\ref{iames},\ref{iseti}}
   \and S.~Rinehart\inst{\ref{inasa}}
   \and S.~Seager\inst{\ref{imit},\ref{imitate},\ref{imiteap}}
   \and P.~Tenenbaum\inst{\ref{iseti},\ref{iames}}
   \and E.B.~Ting\inst{\ref{iames}}
   \and R.~Vanderspek\inst{\ref{imit}}
   \and M.~Vezie\inst{\ref{imit}}
   \and J.N~Winn\inst{\ref{iprinceton}}
    }

  \institute{
  	     Instituto de Astrof\'isica de Canarias (IAC), E-38200 La Laguna, Tenerife, Spain\label{iiac}
  	\and Dept. Astrof\'isica, Universidad de La Laguna (ULL), E-38206 La Laguna, Tenerife, Spain\label{iull}
  	\and Centro de Astrobiologia (CSIC-INTA), Carretera de Ajalvir km 4, 28850 Torrejon de Ardoz, Madrid, Spain\label{csic}
  	\and Department of Astronomy, The University of Tokyo, 7-3-1 Hongo, Bunkyo-ku, Tokyo 113-0033, Japan \label{iutda}
  	\and Astrobiology Center, 2-21-1 Osawa, Mitaka, Tokyo 181-8588, Japan \label{iabc}
  	\and Japan Science and Technology Agency, PRESTO, 7-3-1 Hongo, Bunkyo-ku, Tokyo 113-0033, Japan \label{ijsta} 
  	\and National Astronomical Observatory of Japan, 2-21-1 Osawa, Mitaka, Tokyo 181-8588, Japan \label{inao}
  	\and Department of Earth and Planetary Science, The University of Tokyo, Tokyo, Japan \label{iuteps}
  	\and Rheinisches Institut  f\"ur Umweltforschung an der Universit\"at zu K\"oln, Abteilung Planetenforschung, Aachener Str. 209, 50931 K\"oln, Germany \label{ikoln}
  	\and European Space Agency, ESTEC, Keplerlaan 1, 2201 AZ Noordwijk, The Netherlands \label{iesa}
  	\and  Department of Physics, Kyoto Sangyo University, Kyoto, Japan \label{ikyo}
  	\and Department of Physics and Kavli Institute for Astrophysics and Space Research, Massachusetts Institute of Technology, Cambridge, MA 02139, USA \label{imit}
  	\and Center for Astrophysics ${\rm \mid}$ Harvard {\rm \&} Smithsonian, 60 Garden Street, Cambridge, MA 02138, USA \label{icfa}
  	\and Department of Physics and Astronomy, Vanderbilt University, Nashville, TN 37235, USA \label{ivandy}
  	\and American Association of Variable Star Observers, 49 Bay State Road, Cambridge, MA 02138, USA \label{iaavso}
  	\and Astronomy and Physics, Wesleyan University, Middletown, CT, United States; MIT, Cambridge, MA, United States \label{iags}
  	\and Departamento de Astronom\'ia, Universidad de Chile, Camino El Observatorio 1515, Las Condes, Santiago, Chile. \label{iuc}
    \and NASA Ames Research Center, Moffett Field, CA 94035, USA \label{iames}
  	\and SETI Institute, 189 Bernardo Avenue, Suite 100, Mountain View, CA 94043, USA \label{iseti}
  	\and NASA Goddard Space Flight Center, Greenbelt, MD 20771, USA \label{inasa}
    \and Department of Aeronautics and Astronautics, Massachusetts Institute of Technology, Cambridge, MA 02139, USA \label{imitate}
    \and Department of Earth, Atmospheric, and Planetary Sciences, Massachusetts Institute of Technology, Cambridge, MA 02139, USA \label{imiteap}
  	\and Department of Astrophysical Sciences, Princeton University, Princeton, NJ 08544, USA \label{iprinceton}
  }

  \date{Received September 15, 1996; accepted March 16, 1997}

  \abstract
   {We report the discovery of \tplanet (\tic), a transiting substellar object ($R = \rtmedian\,\rjup$) orbiting a faint M3.5~V dwarf ($V=18.97$) on a 0.56~d orbit. }
   {We set out to determine the nature of the TESS planet candidate \tplanet using ground-based multicolour transit photometry. The host star is faint, which makes RV confirmation challenging, but the large
   transit depth makes the candidate suitable for validation through multicolour photometry.}
   {Our analysis combines three transits observed simultaneously in $r'$, $i'$, and $z_\mathrm{s}$ bands using the MuSCAT2 multicolour imager, three LCOGT-observed transit light curves in $g'$, $r'$, and $i'$ bands, a TESS light curve from Sector 3, and a low-resolution spectrum for stellar characterisation observed with the ALFOSC spectrograph. We model the light curves with \pytransit using a transit model that includes a physics-based light contamination component that allows us to estimate the contamination from unresolved sources from the multicolour photometry. This allows us to derive the true planet-star radius ratio marginalised over the contamination allowed by the photometry, and, combined with the stellar radius, gives us a reliable estimate of the object's absolute radius.}
   {The ground-based photometry strongly excludes contamination from unresolved sources with a significant colour difference to \tstar. Further, contamination from sources of same stellar type as the host is constrained to levels where the true radius ratio posterior has a median of \ktmedian and a 99 percentile of \ktupper. The median and maximum radius ratios correspond to absolute planet radii of \rtmedian and \rtupper~\rjup, respectively, which confirms the substellar nature of the planet candidate. The object is either a giant planet or a brown dwarf (BD) located deep inside the so-called "brown dwarf desert". Both possibilities offer a challenge to current planet/BD formation models and makes \tplanet an object deserving of in-depth follow-up studies.}
   {}

   \maketitle

\section{Introduction}
\label{sec:introduction}

The \emph{Transiting Exoplanet Survey Satellite} (\tess) mission 
is expected to discover thousands of transiting exoplanet candidates orbiting bright nearby stars.
However, since various astrophysical phenomena can lead to a photometric signal
mimicking an exoplanet transit \citep{Cameron2012},
only a fraction of the candidates will be legitimate planets
\citep{Moutou2009,Almenara2009,Santerne2012,Fressin2013},  and the true nature of the
candidates needs to be resolved by
follow-up observations \citep{Cabrera2017a,Mullally2018}. A mass estimate based on radial velocity (RV) measurements offers the most reliable way for candidate confirmation, but RV observations are practical only for bright, slowly-rotating, host stars.  Alternative validation methods need to be applied for candidates around hosts not amenable to RV follow-up.

We report the discovery of \tplanet (\tic), a transiting substellar object ($\rtlower\,\rjup < R < \rtupper\,\rjup$)
orbiting a faint M dwarf ($M_\star=0.4 \pm 0.1\,\msun$, $R_\star=0.405 \pm 0.077\,\rsun$, $V=18.97\pm 0,2$) on a 0.56~d orbit. The object was originally identified in the TESS Sector 3 photometry by the TESS \emph{Science Processing Operations Center} (SPOC) pipeline \citep{Jenkins2016}, and was later followed up from the ground using multicolour transit photometry and low-resolution spectroscopy. The planet candidate passes all the SPOC Data Validation tests \citep{Twicken2018}, and is either a planet or a brown dwarf located in a very sparsely populated region in substellar object period-radius space. 
\begin{table}[t]    
  \caption{\tstar identifiers, coordinates, properties, and magnitudes. The stellar properties
  are based on a spectrum observed with ALFOSC.}
  \centering
  \begin{tabular*}{\columnwidth}{@{\extracolsep{\fill}} llrr}
  \toprule\toprule
  \multicolumn{4}{l}{\emph{Main identifiers}}     \\
  \midrule     
  TIC & &  \multicolumn{2}{r}{120916706} \\
  2MASS        & & \multicolumn{2}{r}{J02282595-2505505}   \\ 
  \\
  \multicolumn{4}{l}{\emph{Equatorial coordinates}}     \\
  \midrule            
  RA \,(J2000) &  & \multicolumn{2}{r}{$2^h\,28^m\,25\fs99$}            \\
  Dec (J2000)  &  & \multicolumn{2}{r}{ $-25\degr\,05\arcmin\,50\farcs39$}  \\
  \\     
  \multicolumn{4}{l}{\emph{Stellar parameters }} \\
  \midrule
  Effective temperature & \teff & [K] & $ 3250 \pm 140$ \\
  Mass    & \smass &[\msun]  & $ 0.4 \pm 0.1 $ \\ 
  Radius  & \srad  &[\rsun]  & $ 0.405 \pm 0.077 $ \\ 
  Age & &[Gyr]&  0.5--9  \\
  Parallax & & [mas] & 3.58\,$\pm$\,0.10\\
  Spectral type & & & M3.5 V\,$\pm$\,0.5 \\
  \\
 \multicolumn{4}{l}{\emph{Magnitudes}} \\
 \midrule              
 \centering

 Filter & & Magnitude       & Uncertainty  \\
 \midrule     
 TESS & & 15.851 & 0.062 \\
 $B$  & & 19.513 & 0.171 \\
 $V$  & & 18.970 & 0.200 \\
 GAIA & & 16.652 & 0.004 \\
 $J$  & & 14.078 & 0.030 \\
 $H$  & & 13.450 & 0.038 \\
 $K$  & & 13.246 & 0.040 \\
  \bottomrule
  \end{tabular*}
  \tablefoot{The \teff estimate is based on \citet{Rajpurohit2013} and \citet{Pecaut2013}, the $R_\star$ estimate on \citet{Schweitzer2019}, and the $M_\star$ estimate on \citet{Maldonado2015}.}
  \label{tbl:star}  
\end{table}

The faintness of \tstar (see Table~\ref{tbl:star}) makes the planet candidate challenging for RV
confirmation.\footnote{Considering the existing instruments, RV follow-up could be feasible 
using 4-VLT mode of ESPRESSO.} However, multicolour transit photometry can be used to validate the nature 
of the candidate \citep{Rosenblatt1971, Drake2003, Tingley2004, Tingley2014a, Parviainen2019}.

Transiting planet candidate validation through multicolour transit photometry works by constraining the 
light contamination from unresolved sources (blending). This allows us to detect blended eclipsing
binaries and, combined with a physics-based light contamination model, allows us to estimate
the transiting object's uncontaminated radius ratio (\emph{true radius ratio}). Combining the true radius
ratio estimate with an estimate of the stellar radius yields the absolute radius of the transiting
object, and if the absolute radius is securely below the theoretical radius limit for a brown dwarf, 
the candidate can be considered a planet.

Our analysis is based on three nights of simultaneous ground-based multicolour transit photometry 
in $r'$, $i'$, and $z_\mathrm{s}$ bands taken with MuSCAT2 multicolour imager \citep{Narita2018} installed in the 
1.5~m Telescopio Carlos Sanchez (TCS) in the Teide Observatory, three transit light curves observed
in $g'$, $r'$, and $i'$ bands with the SINISTRO cameras in the 1~m LCOGT telescopes, and a TESS light 
curve from Sector 3.
The analysis uses a light contamination model included in \pytransit~v2, and yields the posterior 
densities for the model parameters defining the candidate's geometry and orbit, as well as an estimate 
of the true radius ratio in the presence of possible light contamination from blended sources, or 
from the transiting object itself.$\!$\footnote{The effects from a self-illuminating transiting body
contributing flux to the light curve are the same as from a contaminating third body, and are modelled
by the approach without any special modifications.}

The analysis code is available with the data and supplementary material (such as per-dataset analyses 
and posterior sensitivity tests) from GitHub,$\!$\footnote{\giturl} and we encourage anyone interested 
to scrutinise the code (and the underlying assumptions) to ensure its integrity.

\section{Observations}

\subsection{TESS photometry}
\label{sec:observations.tess}

TESS observed \tplanet during Sector 3 for 27~days covering 35 transits with a 2~min cadence. 
We chose to use the Simple Aperture Photometry (SAP) light curves produced by the SPOC 
pipeline \citep{Jenkins2016} over the Presearch Data Conditioning (PDC) light curves,
since the noise in the light curve is dominated by the photon noise (PDC adds some noise but 
did not improve the photometry in this case), and since the PDC process removes the PDC-estimated
flux contamination. The latter can introduce bias into our contamination estimation if the 
PDC contamination is overestimated since we do not allow for 'negative contamination'.

The TESS photometry used in the analysis consists of 35 subsets spanning 2.4~h centred around
each transit based on the linear ephemeris, and each subset was normalised to its median out-of-transit
(OOT) level assuming a transit duration of 0.96~h. The photometry has an average ptp scatter 
of 65~ppt (65000~ppm). We do not detrend the photometry, but include a free baseline level and
white noise standard deviation as per-transit free parameters in the analyses. We also experimented
with higher-order polynomial baseline models and Gaussian Process-based likelihood models, as mentioned
later in Sect.~\ref{sec:analysis}, but these did not change the parameter posteriors due to the 
dominance of photon noise.

\subsection{MuSCAT2 photometry}
\label{sec:observations.muscat2}
We observed two full and one partial transits of \tplanet with the MuSCAT2 multicolour imager \citep{Narita2018} installed in Telescopio Carlos Sanchez (TCS) in the Teide Observatory on the 
nights of 18.12.2018, 19.12.2018, and 02.01.2019. MuSCAT2 is a four-colour instrument consisting
of four independently controlled CCDs, but one of the CCDs was under maintenance, and the observations were carried out simultaneously in three colours ($r'$, $i'$, $z_\mathrm{s}$). 

\begin{figure}
	\centering
	\includegraphics[width=\columnwidth]{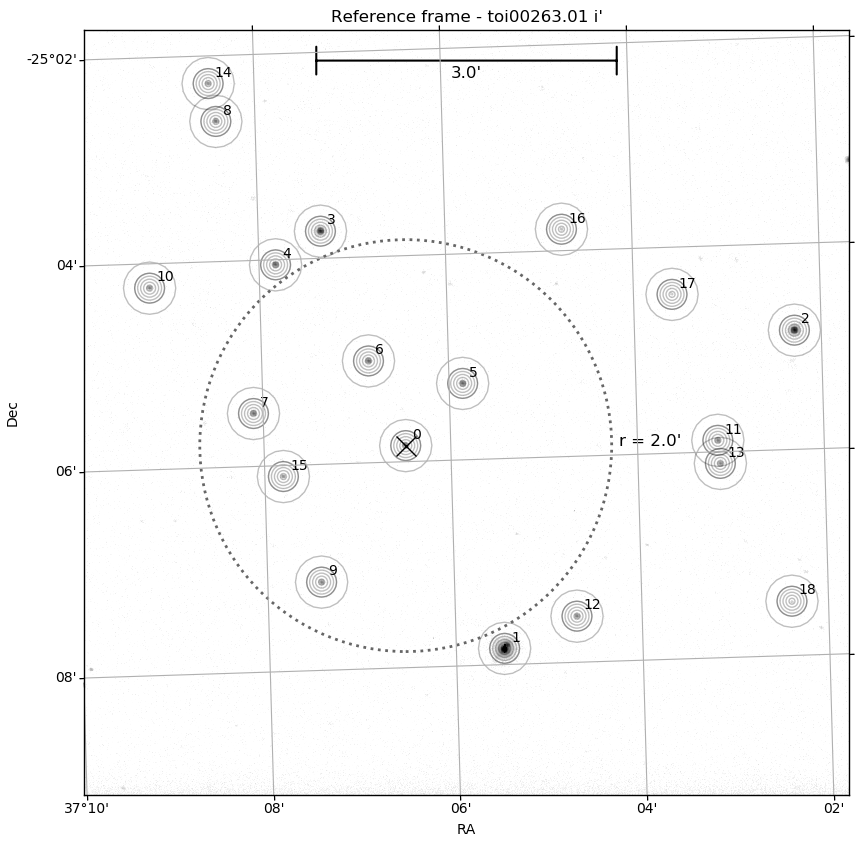}
	\caption{MuSCAT2 field observed in $i'$ band with \tstar marked as 0 and the rest of the stars numbered from brightest to faintest. The circles show the apertures used to extract the photometry, and the two outermost circles mark the annulus used to measure the sky level. The dotted circle
	marks a $2\arcmin$ circle centerd around \tstar.}
	\label{fig:m2_field}
\end{figure}

The exposure times for the first night were 30, 60, and 60~s; for the second night 30, 90, and
90~s; and for the last night 60, 60, and 60~s, with white noise estimates of 47, 14, 13, 47, 14,
12, 13, 8, 8~ppt, respectively. The $r'$-band was used for guiding and had a short exposure time
during the first two nights (which combined with the redness of \tstar explains the
high white noise level). A longer exposure time was used for the guiding channel on the 
third night without any negative effects on the photometry. On the contrary, the last night
had the best observing conditions and the quality of the photometry is significantly higher
than on the first two nights.

The photometry was carried out using standard aperture photometry calibration and reduction steps with a dedicated MuSCAT2 photometry pipeline. The pipeline calculates aperture photometry for a set of comparison
stars and photometry aperture sizes (see Fig.~\ref{fig:m2_field}), and creates the final relative light curves via global optimisation of the
posterior density for a model consisting of a transit model, apertures, comparison stars, and a linear
baseline model with the airmass, seeing, x- and y-centroid shifts, and the sky level as covariates
(that is, the best comparison stars and aperture sizes are also optimised). 

The resulting light curves are stored as fits files with the raw and reduced photometry, best-fitting
transit model, best-fitting baseline model, the covariates, and the covariate coefficients producing 
the baseline model included in binary tables. The MuSCAT2 pipeline reduction also yields posterior
estimates for the transit model, including a contamination estimate based on the approach described 
in Sect.~\ref{sec:transit_model} and detailed further in \citet{Parviainen2019}. While the final
analysis presented in Sect.~\ref{sec:analysis} combines all the photometric data into a joint analysis,
these estimates are important as a consistency check to study the night-to-night variations in the
parameters of interest.

\subsection{LCOGT photometry}
\label{sec:observations.lco}

Three full transits of \tplanet were observed using the Las Cumbres Observatory Global Telescope (LCOGT) 1~m network \citep{Brown2013} in $i'$, $r'$, and $g'$ bands on the nights of 13.12.2018, 26.12.2018, and 21.01.2019, respectively, as part of the TESS Follow-up Observing Program (TFOP). We used the {\tt TESS Transit Finder}, which is a customised version of the {\tt Tapir} software package \citep{Jensen2013}, to schedule our transit observations. The $g'$ and $i'$ transits were observed from the LCOGT node at South Africa Astronomical Observatory and used 200~s and 100~s exposures respectively. The $r'$ transit was observed from the LCOGT node at Cerro Tololo Inter-American Observatory and used 120~s exposures. The telescopes are equipped with $4096\times4096$ LCO SINISTRO cameras having an image scale of 0$\farcs$389 pixel$^{-1}$ resulting in a $26\arcmin\times26\arcmin$ field of view. 

The images were calibrated by the standard LCOGT BANZAI pipeline and the photometric data were extracted using the {\tt AstroImageJ} ({\tt AIJ}) software package \citep{Collins2017}. Circular apertures with radius 5, 8, and 8 pixels were used to extract differential photometry in the $g'$, $r'$, and $i'$-bands resulting in estimated white noise of 28-ppt, 7.55~ppt, and 7.55~ppt, respectively. The images have stellar point-spread-functions (PSFs) with FWHM ranging from $1\farcs2$ to $1\farcs8$. The nearest star in the \gaia DR2 catalogue is $21\farcs2$ to the north of \tstar, so the photometric apertures are not contaminated with significant flux from known nearby stars.

\subsection{ALFOSC spectroscopy}
\label{sec:observations.alfosc}

We used low-resolution optical spectrum observed using the Alhambra Faint Object Spectrograph and Camera (ALFOSC) of the 2.5-m Nordic Optical Telescope (NOT) on Roque de los Muchachos Observatory (La Palma) to obtain the basic
stellar properties of \tstar. A total of two spectra with on-source exposure times of 1800~s each were obtained on 2019 January 21. ALFOSC is equipped with an E2V 2048\,$\times$\,2048 CCD detector with a pixel size of 0\farcs2138 projected onto the sky. We used the grism number 5 and a long slit width of 1\farcs0, which yield spectra between 5000 and 9200 \AA~with spectral resolution of 16.6 \AA~($R = 430$ at 7100 \AA). Seeing conditions and sky transparency were fine for carrying out spectroscopic observations. \tstar was observed at parallactic angle and at a relatively low airmass of 1.7. 

ALFOSC raw frames were reduced following standard procedures at optical wavelengths: bias subtraction, flat-fielding using dome flats, and optimal extraction using appropriate packages within the IRAF\footnote{Image Reduction and Analysis Facility (IRAF) is distributed by National Optical Astronomy Observatories, which is operated by the Association of Universities for Research in Astronomy, Inc., under contract with the National Science Foundation.} environment. Wavelength calibration was performed using He\,{\sc i} and Ne\,{\sc i} arc lines observed immediately after the acquisition of the target data. The instrumental response was corrected using observations of the spectrophotometric standard star GJ\,246 (a white dwarf) taken with exactly the same instrumental configuration as our target on 2019 February 4. Unfortunately, the standard was observed at the low airmass of 1.2; we cannot use it for removing the telluric contribution from the target data. The two individual spectra of \tstar~were combined and the final spectrum, depicted in Figure~\ref{alfosc_spectrum}, has very good quality (S/N > 150) for a proper spectral classification. 

\section{Stellar characterisation}
\label{sec:observations.alfosc_details}

\begin{figure*}
	\centering
	\includegraphics[width=\textwidth]{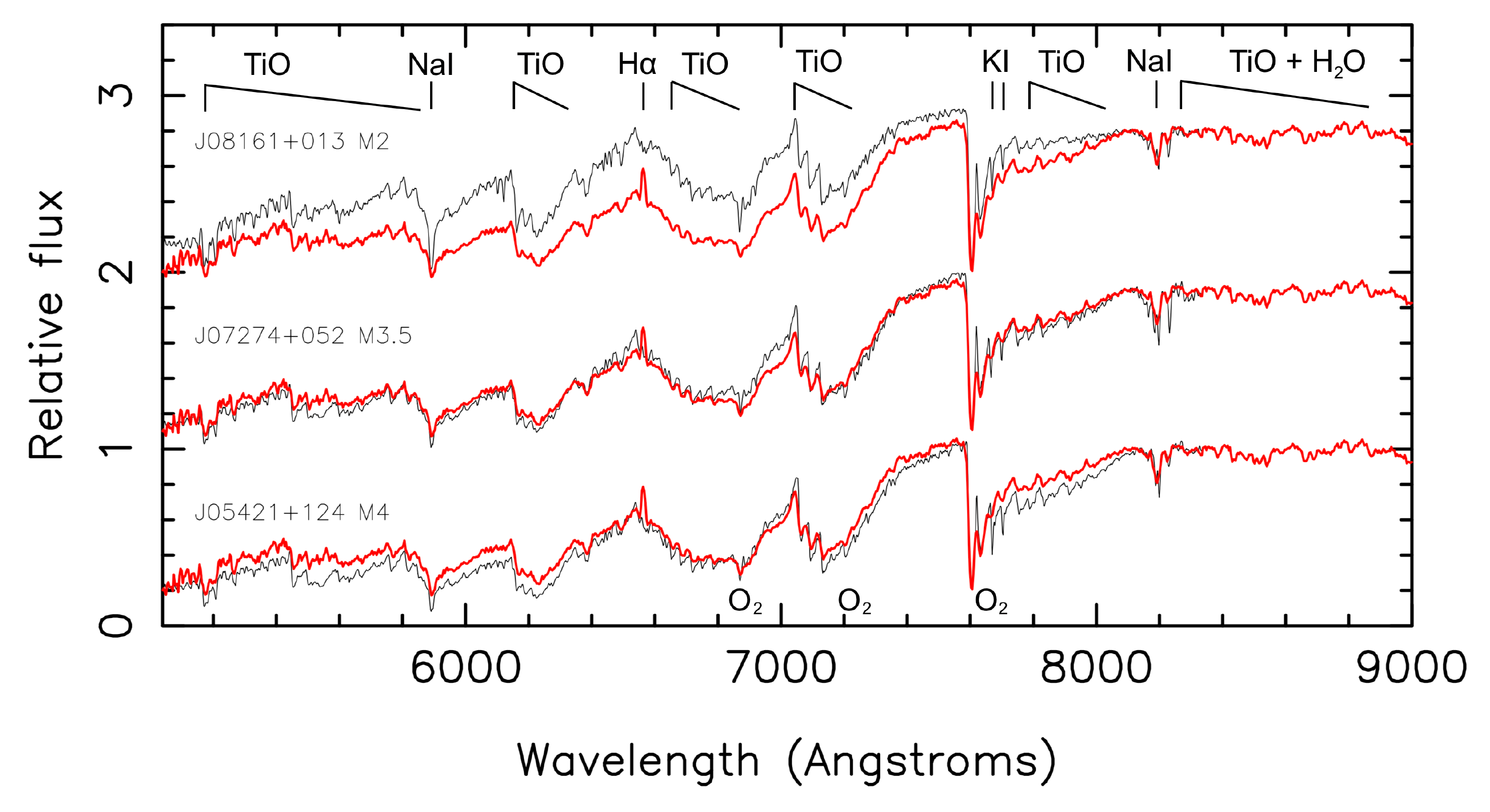}
	\caption{ALFOSC spectrum of TOI 263 with a resolution of 16.6 \AA~is plotted as a red line. It is compared to the spectra of three spectral standard stars from the catalogue of Alonso-Floriano et al. (2015), which are shown with a black line: GJ\,2066 (J08161$+$013, M2V), Luyten’s star (J07274$+$052, M3.5V), and V1352\,Ori (J05421$+$124, M4V). All spectra are normalised to 1.0 at the red continuum of the subordinate Na\,{\sc i} lines, and are shifted vertically by 1.0 for clarity. The strongest molecular and atomic features are indicated on the top. The strongest telluric oxygen features are also labelled.}
	\label{alfosc_spectrum}
\end{figure*}

\tstar shows strong absorption due to TiO all over the ALFOSC spectrum while VO is not present, and the observed  pseudo-continuum increases toward long wavelengths, all of which clearly indicates an early- to mid-M spectral type. We employed the spectroscopic standard stars defined in Table~3 by \citet{alonso-floriano15} to derive an M3.5~V spectral type for \tstar, since Luyten's star (also depicted in Figure~\ref{alfosc_spectrum}) provides the best match to the ALFOSC data. \citet{alonso-floriano15} spectra have slightly higher resolution by a factor of 3.5 than our data, however, the comparison is feasible given the overlapping wavelength coverage. The different spectral resolution does not affect the spectral typing of \tstar. We estimated an error of $\pm$0.5 subtypes for our classification. \tstar shows H$\alpha$ in emission with a pseudo-equivalent width (pEW) of $-3.9 \pm 0.2$ \AA. This width is typical among field M3--M4-type stars (see Figure~8 of \citealt{alonso-floriano15}) and lies well below the threshold defined for accreting M dwarfs (see \citealt{barrado03}), thus indicating that \tstar is quite likely not a very young star in the solar neighbourhood. Given the low spectral resolution of our data, we could not assess the star's surface gravity with precision; however, the Na\,{\sc i} resonance doublet at $\sim$8195 \AA, which is a well-known age-indicator for M dwarfs, is clearly detected with pEW = $3.6 \pm 0.2$ \AA{} in \tstar. This value is typical of field, high-gravity M3--M4 dwarfs (see Table~5 by \citealt{martin96}, see also \citealt{schlieder12}), which suggests that \tstar likely has the "age of the field", i.e., between $\approx$0.5 and $\approx$9 Gyr. Similarly to the surface gravity, the star's metallicity cannot be addressed quantitatively using the ALFOSC data. However, there is no strong absorption due to hydrides in the observed optical spectrum, thus hinting that \tstar likely has a near solar chemical composition {\color{red}\citep[][and the references therein]{Kirkpatrick2014}}.

\section{Transit Light curve analysis}
\label{sec:analysis}

\subsection{Overview}
We first analysed each photometric dataset (MuSCAT2, TESS, and LCOGT) independently, and then 
combined them for a joint analysis. We do not detail the independent analyses here, but make them 
available online as Jupyter notebooks in GitHub.

The final dataset consists of the 35 transits in the TESS data from Sector 3, three transits observed
simultaneously in three passbands with MuSCAT2, and three transits observed in three passbands with 
the LCOGT telescopes. This sums up to five passbands, 41 transits, and 47 light curves.

The analysis follows standard steps for Bayesian parameter estimation \citep{Parviainen2018}. First, 
we construct a flux model 
that aims to reproduce the transit and the light curve systematics. Next, we define a noise model to
explain the stochastic variability in the observations not explained by the deterministic flux model.
Combining the flux model, the noise model, and the observations gives us the likelihood.  Finally, we
define the priors on the model parameters, after which we estimate the joint parameter posterior
distribution using Markov Chain Monte Carlo (MCMC) sampling.

The posterior estimation begins with a global optimisation run using Differential Evolution
\citep{Storn1997,Price2005} that results with a 
population of parameter vectors clumped close to the global posterior mode. This parameter vector 
population is then used as a starting population for the MCMC sampling with \textsc{emcee}, and the
sampling is carried out until a suitable posterior sample has been obtained \citep{Parviainen2018}.

The analyses were carried out with a custom Python code based on \pytransit~v2\footnote{\url{https://github.com/hpparvi/pytransit}} \citep{Parviainen2015,Parviainen2019}, which includes a physics-based contamination model based 
on the \textsc{PHOENIX}-calculated stellar spectrum library by \citet{Husser2013}. The limb darkening computations were carried 
out with \ldtk$\!$\footnote{\url{https://github.com/hpparvi/ldtk}} \citep{Parviainen2015b}, and 
MCMC sampling was carried out with \textsc{emcee} \citep{Foreman-Mackey2012,Goodman2010}.

The code relies on the existing \textsc{Python} packages for scientific computing and astrophysics:
\textsc{SciPy}, \textsc{NumPy} \citep{VanderWalt2011}, \textsc{AstroPy}
\citep{TheAstropyCollaboration2013}, \textsc{photutils} \citep{Bradley2019}, \textsc{astrometry.net} \citep{Lang2010}, \textsc{IPython} \citep{Perez2007},
\textsc{Pandas} \citep{Mckinney2010}, \textsc{xarray} \citep{Hoyer2017}, \textsc{matplotlib} \citep{Hunter2007}, and
\textsc{seaborn}. 

All the code for the analyses presented in this paper is available from GitHub as Python code and 
Jupyter notebooks from \giturl.

\subsection{Contaminated transit model}
\label{sec:transit_model}

The candidate validation is based on the \emph{true planetary radius} estimate obtained modelling the
multicolour transit photometry with a transit model that includes a physics-based light contamination 
component. The contamination is parametrised by the effective temperatures of the host and the contaminant(s) 
(\teffh and \teffc, respectively), and the amount of contamination in some reference passband (\pcref)
\citep{Parviainen2019}. 

The apparent area and radius ratios ($k^2_\mathrm{app}$ and \kapp, respectively) can be directly estimated
from the transit light curve, and are related to the true radius ratio, \ktrue, as
\begin{equation}
    \kapp \sim \sqrt{\Delta F} = \ktrue \sqrt{1-c}, \label{eq:kapp}
\end{equation}
where $\Delta F$ is the transit depth, and \pcref is the passband dependent contamination factor. The true 
radius ratio is derived from \kapp then as
\begin{equation}
    \ktrue = R_\mathrm{p}/R_\star = \kapp / \sqrt{1-\pcref}. \label{eq:ktrue}
\end{equation}
The contamination factor depends on the wavelength if the host and contaminant(s) are  of different spectral
type (that is, if $\teffh \neq \teffc$), which leads
to passband dependent variations in the transit depth ($k^2_\mathrm{app}$). Besides the transit depth variations,
a subtle colour dependent signature exists that is directly related to the true radius ratio,
and constrains the contamination even if $\teffh = \teffc$
\citep{Drake2003,Tingley2004,Tingley2014a,Parviainen2019}.

The contamination model yields a contamination estimate for the observed passbands given \teffh, \teffc,
\pcref, and \kapp, where the reference passband for \pcref and \kapp is freely chosen (and does not affect
the posterior estimate). The contaminated transit model for the passband $i$ is now
\begin{equation}
    \tmodel_{c,i} = c_i + (1-c_i) \times \tmodel,
\end{equation}
where \tmodel is the uncontaminated transit model.

The final \ktrue estimate is marginalised over the contamination allowed by the photometry (and all the other
model parameters),  including contamination from sources of similar spectral type as the host 
star.$\!$\footnote{The analysis is valid also
in the case of a self-illuminating transiting object, such as in the case of a eclipsing binary system.}

\subsection{Log posterior}

The log posterior for a parameter vector \pvec given a combined dataset $\vec{D}$ with $n_\mathrm{t}$ transits observed in $n_\mathrm{b}$ unique passbands is
\begin{equation}
 \ln P(\pvec|\vec{D}) = \ln P(\pvec)
                + \sum_i^{n_\mathrm{t}} \llh(\vec{D}_\mathrm{i}|\pvec)
                + \sum_j^{n_\mathrm{b}} \llh_\mathrm{ld}(\vec{\phi}_j),
\end{equation}
where the first term is the log prior, the second is the total log likelihood, the last 
term is the sum of the \ldtk-calculated log likelihoods for the limb-darkening (when using \ldtk to
constrain the stellar limb darkening, as explained below), and $\vec{\phi}_j$ is a subset of \pvec
containing the limb darkening coefficients for the $j$th passband.  

\subsection{Noise model and log likelihood}

In general, the log likelihood for a single transit light curve assuming normally distributed noise is
\begin{equation}
 \llh(\vec{D}|\pvec) = -\frac{1}{2} \left( n_\mathrm{D} \ln 2\pi +\ln|\covmat| +\vec{r}^\mathrm{T} 
\covmat^{-1} 
\vec{r}\right),
 \label{eq:lnlikelihood_gn}
\end{equation}
where $n_\mathrm{D}$ is the number of datapoints, $\vec{r}$ is the residual vector 
($\vec{F}_o - F(\pvec|\vec{t},\vec{C})$), and $\covmat$ is the covariance matrix.

As already mentioned, we experimented using Gaussian processes to model the residual correlated noise 
not explained by the linear baseline model described below, but chose to simplify our approach after 
tests against uncorrelated noise model did not show significant differences. In the end, we chose 
a normally distributed uncorrelated noise model that leads to the standard likelihood equation
\begin{equation}
  \ln P(\vec{D}|\pvec,\sigma) = -\frac{1}{2}\left(n\ln2\pi +\sum_i^n \ln \sigma_{\mathrm{i}}^2 + \sum_{i=1}^n \frac{r_i^2}{2\sigma_{\mathrm{i}}^2} \right ), 
\end{equation}
where $\sigma$ are the per-point photometric errors. Further, our photometric errors do not show great
variations within individual transits, so we chose to use a log average per-transit photometric error as 
a free parameter in the model.

\subsection{Light curve model}

The flux for a single observation $i$ is modelled as
\begin{equation}
    F(\pvec|t_i,\vec{c}_i) = \tmodel_\mathrm{c}(\pvec|t_i) \times  B(\pvec|\vec{c}_i),
\end{equation}
where $\tmodel_\mathrm{c}$ is the (contaminated) transit model, $B$ is the baseline model, \pvec is the parameter
vector, $t$ is the mid-exposure time, and $\vec{c}_i$ is the covariate vector.

We use the standard quadratic Mandel~\&~Agol transit model \citep{Mandel2002} implemented in 
\pytransit with the triangular parametrisation presented by \citet{Kipping2013b}.

We use a simple linear model to represent the baseline flux variations as a function of linear
combination of light-curve-specific set of covariates. The linear model for the final joint analysis
was chosen after carrying out analyses for the individual MuSCAT2 and LCO datasets using Gaussian 
processes (GPs, \citealt{Rasmussen2006,Gibson2011a,Roberts2013}) to model the systematics. The GP
and linear model analyses agreed with each other, and we chose the computationally faster approach
for the final analysis.

\subsection{Priors and model parametrisation}
\subsubsection{Joint model parametrisation}

\begin{table}[t]
 \caption{Final joint model parametrisation and priors. \UP{a,b} stands for a uniform prior from $a$  
 to $b$, and \NP{\mu,\sigma} for a normal prior with mean $\mu$ and standard deviation $\sigma$. The
 division of parameters into different categories is explained in the main text.}
 \centering
 \begin{tabular*}{\columnwidth}{@{\extracolsep{\fill}}lll}
  \toprule\toprule
  Notation & Name & Prior \\
  \midrule
  \multicolumn{3}{l}{\textit{System parameters}} \\
  $T_c$  & zero epoch & $\NP{2458386.1723, 0.0015} $\\
  $P$    & orbital period & $\NP{0.5567365, 1\times 10^{-5}} $ \\
  $\rho_\star$ & stellar density & $\UP{0.10, 25} $ \\
  $b$ & impact parameter & $\UP{0, 1} $ \\
  $k^2_\mathrm{true}$ & True area ratio & $\mathcal{U}(0.10^2, 0.75^2)$ \\
  \\
  \multicolumn{3}{l}{\textit{Passband independent contamination parameters}} \\
  \teffh & host \teff & $\NP{3116, 100}$ \\
  \teffc & contaminant \teff & $\UP{2500, 12000}$ \\ 
  \\
  \multicolumn{3}{l}{\textit{Passband dependent contamination parameters}} \\
  $k^2_\mathrm{app}$ & apparent area ratio& $\mathcal{U}(0.1^2, 0.3^2)$ \\
  \\
  \multicolumn{3}{l}{\textit{Stellar limb darkening, passband dependent}} \\
  $q_1$ & limb darkening q$_1$ & \UP{0,1} or \ldtk \\
  $q_2$ & limb darkening q$_2$ & \UP{0,1} or \ldtk \\
  \\
  \multicolumn{3}{l}{\textit{Average photometric error, light curve dependent}} \\
  $\log_{10}{\sigma}$ & $\log_{10}$ error  & $\UP{-4, 0}$ \\
  \\
  \multicolumn{3}{l}{\textit{TESS baseline coefficients, light curve dependent}} \\
  $c_i$ & intercept & $\NP{1, \sigma_F}$ \\
  \\
  \multicolumn{3}{l}{\textit{MuSCAT2 baseline coefficients, light curve dependent}} \\
    $c_i$ & intercept & $\NP{1, \sigma_F}$ \\
    $c_b$ & sky & $\NP{0, \sigma_F}$ \\
    $c_b$ & centroid x shift & $\NP{0, \sigma_F}$ \\
    $c_b$ & centroid y shift & $\NP{0, \sigma_F}$ \\
    $c_b$ & aperture entropy & $\NP{0, \sigma_F}$ \\
  \\
  \multicolumn{3}{l}{\textit{LCOGT baseline coefficients, light curve dependent}} \\
    $c_i$ & intercept & $\NP{1, \sigma_F}$ \\
    $c_b$ & airmass & $\NP{0, \sigma_F}$ \\
    $c_b$ & centroid x shift & $\NP{0, \sigma_F}$ \\
    $c_b$ & centroid y shift & $\NP{0, \sigma_F}$ \\
    $c_b$ & FWHM & $\NP{0, \sigma_F}$ \\
  \bottomrule
 \end{tabular*}
 \label{tbl:parametrisation}
\end{table}

The final joint model needs to reproduce the 47 light curves observed in five passbands with three
instruments. This leads to a somewhat complex parametrisation with 160 free parameters, of which only
5-9 are of physical interest.

We divide the model parameters into three main categories. First come the \emph{system parameters} 
that define the planet candidate's orbit and geometry. These are the quantities that we are actually
interested about, and are directly connected to the physical properties of the planet candidate. The system
parameters do not depend on the observation passband, instrument, or any other external factor,
so all the observations contribute to their likelihood. Next come the \emph{passband dependent}
parameters such as the stellar limb darkening coefficients and apparent planet-star area ratios. 
These are quantities that vary as a function of wavelength, and so change from passband to passband,
but do not depend on the instrument, observing conditions, etc. All the observations carried out
in a given passband contribute to the likelihoods of the passband-dependent parameters in that
passband. Finally come the \emph{light curve dependent} parameters that affect the model of
each independent light curve, such as the linear baseline model coefficients. These are unique
to each light curve (and instrument) and their posteriors are of little practical interest.
However, these parameters need to be included as nuisance parameters to be marginalised over
so that their effect on the model is correctly reflected in the uncertainties of the main parameters 
of interest.

The final parametrisation is outlined in Table~\ref{tbl:parametrisation}. The passband dependent
parameters are repeated for each passband, and the light curve dependent parameters are repeated
for each light curve.

\subsubsection{Baseline coefficients}

TESS, LCOGT, and MuSCAT2 have all different sets of covariates for their baseline models. With TESS we 
include only a constant intercept (the out-of-transit flux level) as a free parameter for each
transit. We also carried out analyses with more complex polynomial baseline models and Gaussian
processes, but the higher-order polynomial coefficient posteriors agreed with zero, and the GP 
hyperparameters favoured models dominated by white noise. With MuSCAT2 we include the intercept,
median sky level, x- and y- centroid shifts, and aperture entropy (a proxy for FWHM). With LCOGT
we include the intercept, airmass, centroid shifts, and FWHM. The intercepts have normal priors
centred to unity and standard deviations set to the standard deviation of per-transit flux
($\sigma_\mathrm{F}$). The rest of the baseline coefficients have normal priors centred around zero
with their standard deviations set as for the intercepts.

\subsubsection{Stellar limb darkening}
\label{sec:analysis.parameters.limb_darkening}
Stellar limb darkening is degenerate with the impact parameter and apparent area ratio, and so
any assumptions taken about limb darkening can bias the estimates of these two parameters \citep{Csizmadia2013,Espinoza2015,Parviainen2018,Parviainen2019}. 
While multicolour transit photometry can break this degeneracy (limb darkening is passband dependent but
the impact parameter and radius ratio are not), we were still interested in understanding how sensitive
our analysis results are on our prior assumptions about limb darkening. 

For that reason, we repeated all the analyses for two cases. First, we carried out the analyses with a
uniform prior on the triangular quadratic limb darkening coefficients from zero to unity \citep[leading 
to uniform priors covering the physically plausible parameter space of the quadratic
coefficients,][]{Kipping2013b}. Next, we repeated the analyses using \ldtk
to constrain the shape of allowed limb darkening profiles \citep[that is, setting a prior in the limb 
darkening profile space instead of setting priors on the coefficients themselves,][]{Parviainen2015b}.

\subsubsection{Contamination}
\label{sec:analysis.parameters.contamination}

The ground-based photometry has been observed with significantly smaller aperture radii than the TESS
photometry, and TESS photometry can a priori be expected to have a different amount of contamination
from third light sources than the ground-based photometry. This breaks the prior assumption for the 
physics-based contamination model described in Sect.~\ref{sec:transit_model} that the contaminating
sources are constant for all the observations. However, the MuSCAT2 and LCOGT photometry was carried 
out with similar-sized apertures, and the assumption can be considered to hold.

Thus, we parametrise the model with two contamination factors. The MuSCAT2 and LCOGT transit models 
use the physics-based contamination model, but the TESS transit model is given an independent 
unconstrained contamination factor. Further, we do not parametrise the model directly with the
true area ratio and contamination, since this was observed to lead to a parameter space that is 
difficult to sample efficiently with MCMC. We use the apparent and true area ratios instead, where
the per-passband contamination levels can be derived using Eqs.~\eqref{eq:kapp} and \eqref{eq:ktrue}.

\section{Results}
\label{sec:results}

\begin{figure*}
	\centering
	\includegraphics[width=\textwidth]{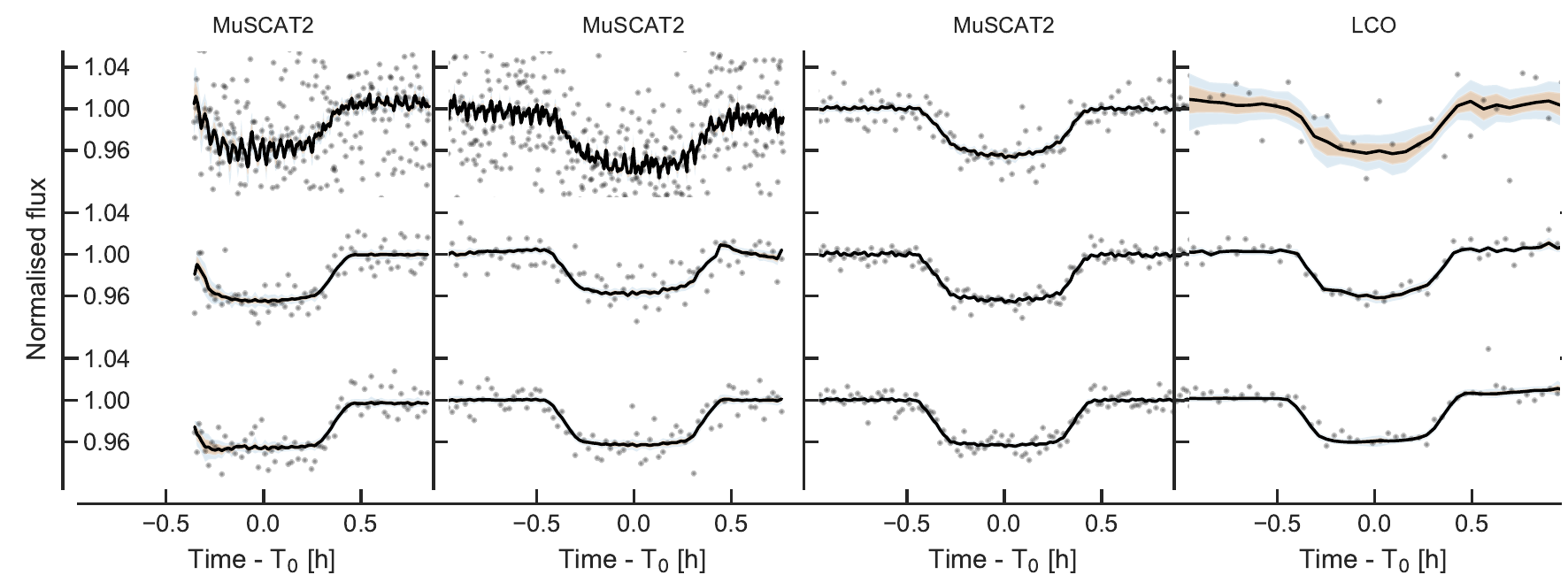}
	\caption{Three transits of \tplanet observed simultaneously in $r'$, $i'$, and $z_\mathrm{s}$ with MuSCAT2, and three separate transits observed with SINISTRO. Each column shows a separate observing night and each row a separate filter for MuSCAT2 observations, while each column show a separate transit observed in
	a separate filter for the LCO data. MuSCAT2 photometry is shown with the original cadences (light grey points) and binned into 4~min bins shown as black points with errorbars showing the standard error of the mean. The LCOGT observations are not binned due to the longer exposure times. The median baseline model has been subtracted from the photometry, and the black lines show the median posterior transit model.}
	\label{fig:lc_m2_lco}
\end{figure*}

\begin{figure}
	\centering
	\includegraphics[width=\columnwidth]{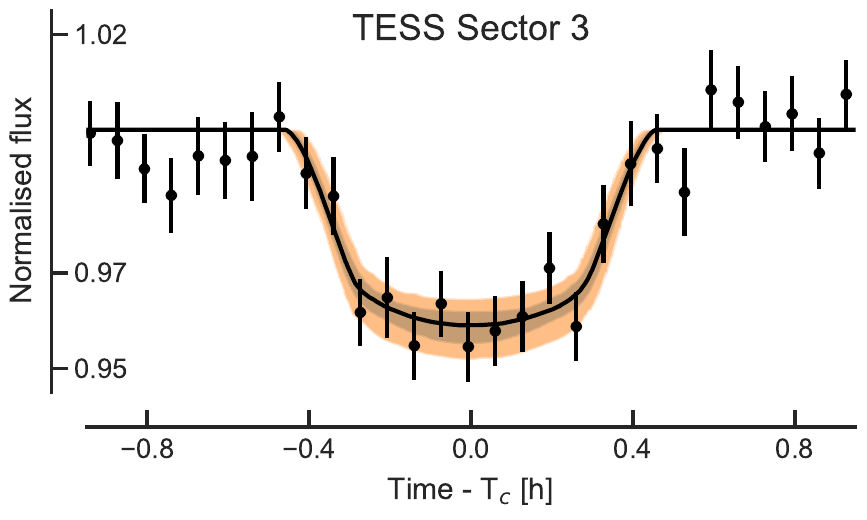}
	\caption{Phase folded and binned TESS Sector 3 light curve (points with uncertainties) and the median transit posterior model with its 16 and 84 percentile limits. The model corresponds to the final joint fit with TESS, MuSCAT2 and LCO observations. The data has been divided by the median baseline model, phase folded, and binned in 4~min bins for visualisation purposes.}
	\label{fig:lc_tess_folded}
\end{figure}

\begin{figure*}
	\centering
	\includegraphics[width=\textwidth]{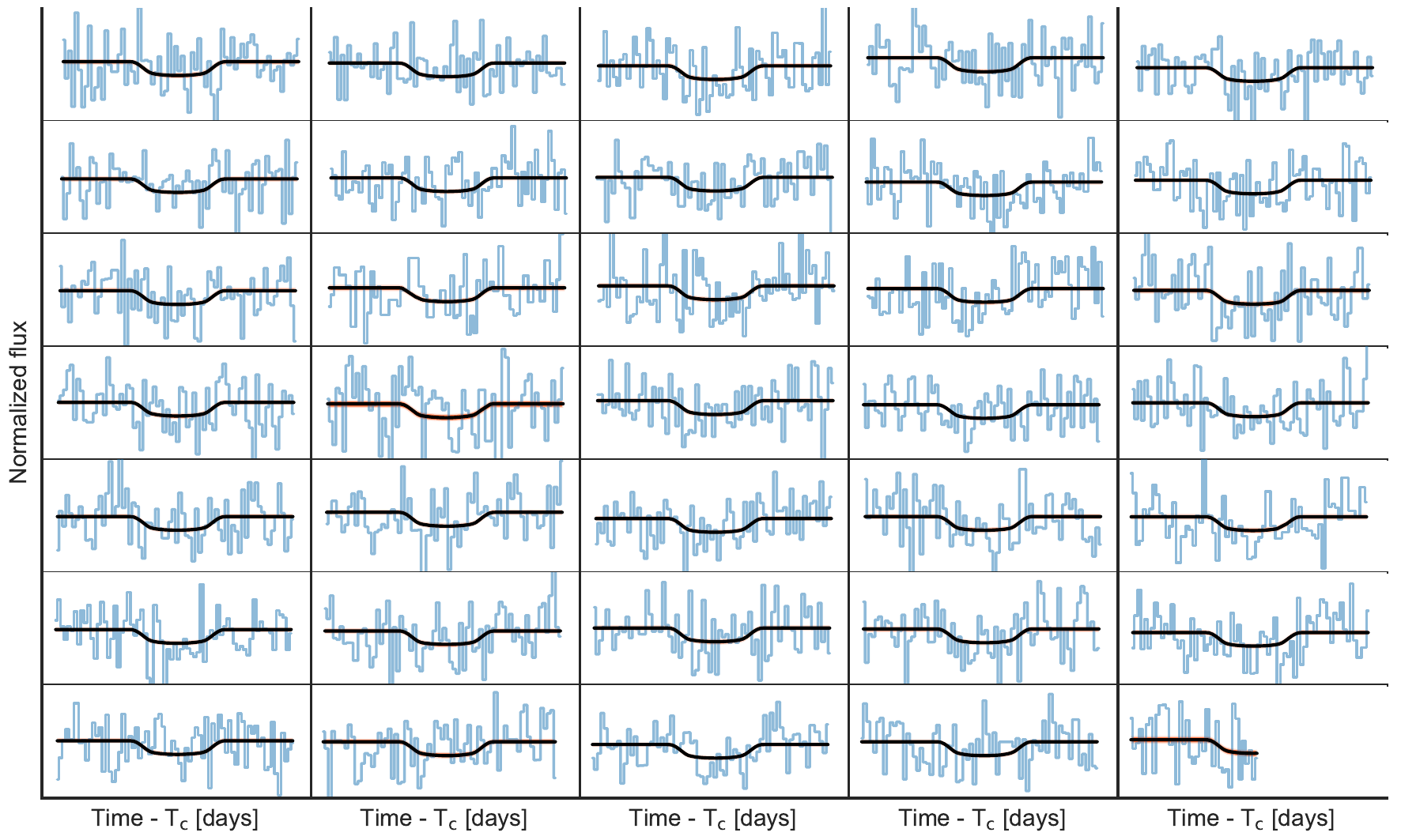}
	\caption{Individual TESS transits with the median posterior transit model and its 68\% central posterior limits.}
	\label{fig:lc_tess_transits}
\end{figure*}

We show the ground-based photometry with the transit model in Fig.~\ref{fig:lc_m2_lco}, the phase-folded 
TESS photometry and model in Fig.~\ref{fig:lc_tess_folded}, and the per-transit TESS photometry and model in
fig.~\ref{fig:lc_tess_transits}.

\begin{figure*}
	\centering
	\includegraphics[width=\textwidth]{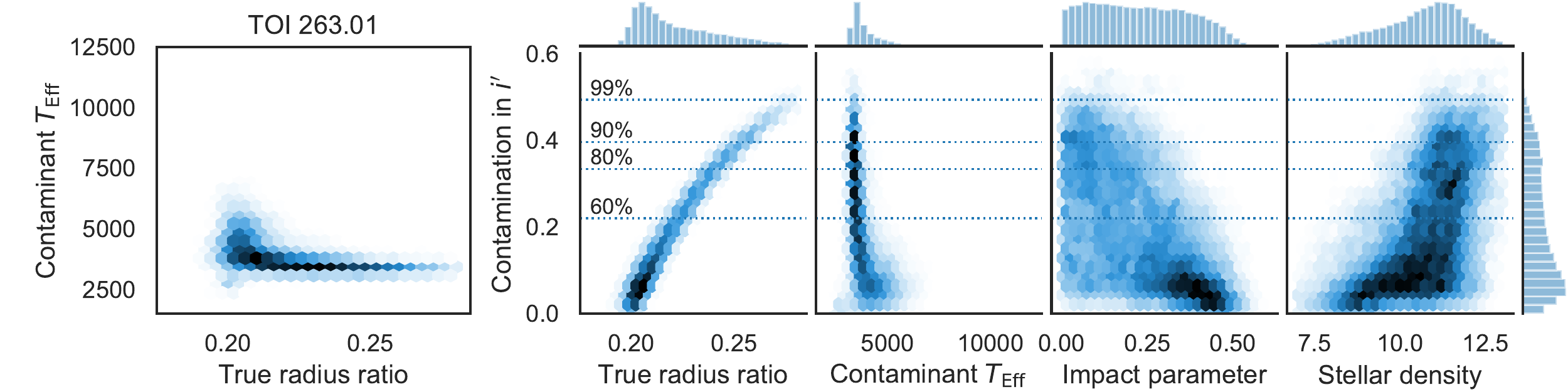}
	\caption{Joint and marginal posterior distributions for the key parameters from the modelling of the TESS, MusCAT2, and LCO transit light curves together.}
	\label{fig:contamination_posteriors}
\end{figure*}

The joint multicolour transit modelling excludes significant levels of blending from sources with effective temperature ($\teffc$) different from that of the host star ($\teffh$), and also strongly constrains allowed blending from sources with $\teffc \sim \teffh$, as shown in Fig.~\ref{fig:contamination_posteriors}. Grazing transit geometries are also excluded, as the impact parameter is constrained to $\pimp < 0.51$ at $99\%$ level. The stellar density posterior median of $11$~\gcm  agrees well with theoretical expectations for an M dwarf with $\teff\approx3200$~K.

\begin{figure}
	\centering
	\includegraphics[width=\columnwidth]{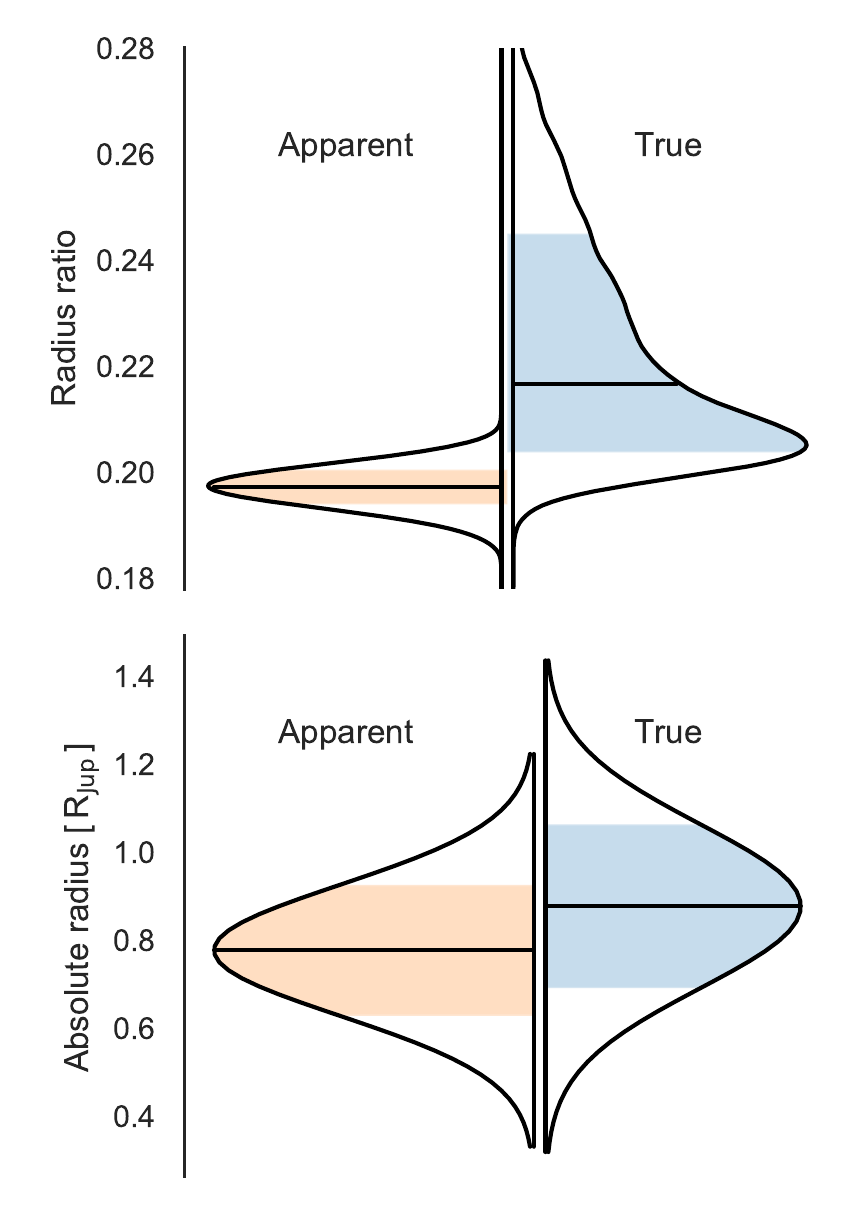}
	\caption{The apparent and true radius ratio posteriors (above), and the apparent and true candidate
	radius posteriors (below). Allowing for blending creates a tail towards high radius ratios, in this
	case corresponding to possible blending with $\teffh \sim \teffc$, but this has only a minor effect
	on the absolute planet radius posterior, which is dominated by the uncertainty in the stellar radius.}
	\label{fig:k_and_rp_posteriors}
\end{figure}

The median posterior value for the true radius ratio, \ktrue, is \ktmedian, with a $99$ posterior percentile 
value of \ktupper. This leads to absolute planetary radius, \rptrue, posterior median of $\rtmedian\,\rjup$ 
with a $99$ posterior percentile limit of $\rtupper\,\rjup$. Thus, the candidate radius is in the size range
shared both by gas giants and brown dwarfs \citep{Burrows2011,Chabrier2000}. Unfortunately, uncertainty in 
\rptrue estimate is dominated by the uncertainty in the stellar radius, as illustrated in
Fig.~\ref{fig:k_and_rp_posteriors}, and further transit observations would not be able to significantly
reduce the uncertainty even if the uncertainty in contamination would be brought to zero.

The analysis takes into account the possibility that the transiting object is bright enough to contribute 
to the total flux significantly. This would be the case of an eclipsing binary system, possibly with
close-to identical components. However, our results exclude the possibility that the transiting body would
be self-illuminating.

We also carried out a TTV search from the TESS light curves using PyTV (Korth, in prep.), but the S/N
ratio of the individual transits was too low for meaningful transit centre estimates.

\begin{table*}
\centering
\small
\caption{Relative and absolute estimates for the stellar and planetary parameters derived from the multicolour transit analysis.}
\begin{tabular*}{\textwidth}{@{\extracolsep{\fill}} llll}        
\toprule\toprule
\multicolumn{4}{l}{\emph{Ephemeris}} \\
\midrule
Transit epoch & $T_0$ & [BJD] & $ 2458386.17184 \pm 6.7 \times 10^{-4}$\\
Orbital period & $P$ & [days] &  $0.5568140 \pm 4.1 \times 10^{-6}$ \\
Transit duration & $T_{14}$ & [h] & $0.912 \pm 0.017$ \\
\\
\multicolumn{4}{l}{\emph{Relative properties}} \\
\midrule
Apparent radius ratio &$k_\mathrm{app}$ & $[R_\star]$ & $ 0.1982 \pm 0.0035$ \\
True radius ratio &$k_\mathrm{true}$ & $[R_\star]$ & $0.2073 \; (-0.0075) \; (+0.0215)$ \\
Scaled semi-major axis &$a_\mathrm{s}$ & $[R_\star]$ & $5.54 \pm 0.22$ \\
Impact parameter &$b$ && $0.29 -0.17 +0.12$ \\
\\
\multicolumn{4}{l}{\emph{Absolute properties}} \\
\midrule 
Apparent companion radius$^a$ &$R_{\mathrm{p,app}}$ & [\rjup]  &  $ 0.78 \pm 0.15 $ \\
True companion radius$^a$ &$R_{\mathrm{p,true}}$ & [\rjup]  &  $ 0.83 \pm 0.17 $ \\
Semi-major axis$^a$& $a$ &[AU] & $ 0.010 \pm 0.002 $\\
Eq. temperature$^b$ & $T_{\mathrm{eq}}$ &[K] & $ 1020 \pm 80 $ \\
Stellar density & $\rho_\star$ & $[\gcm]$ & $10.4 \pm 1.2 $\\
Inclination & $i$ &[deg] & $86.96 \pm 1.6$ \\
\bottomrule       
\end{tabular*}
\tablefoot{ The estimates
correspond to the posterior median ($P_{50}$) with $1 \sigma$ uncertainty estimate based on the 16th
and 84th posterior percentiles ($P_{16}$ and $P_{84}$, respectively) for symmetric, 
approximately normal posteriors. For asymmetric, unimodal, posteriors, the estimates are $P_{50}{}^{P_{84}-P_{50}}_{P_{16}-P_{50}}$. 
.\tablefoottext{a}{The semi-major axis and planet candidate radius are based on the scaled
semi-major axis and true radius ratio samples, and the stellar radius estimate shown in Table~\ref{tbl:star}.} \tablefoottext{b}{The equilibrium temperature of the planet candidate is calculated using the stellar \teff estimate, scaled semi-major axis distribution, heat redistribution factor distributed uniformly between 0.25 and 0.5, and planetary albedo distributed uniformly between 0 and 0.4.}}
\label{table:parameters}  
\end{table*}

\section{Discussion}
\label{sec:discussion}

\begin{figure*}
	\centering
	\includegraphics[width=\textwidth]{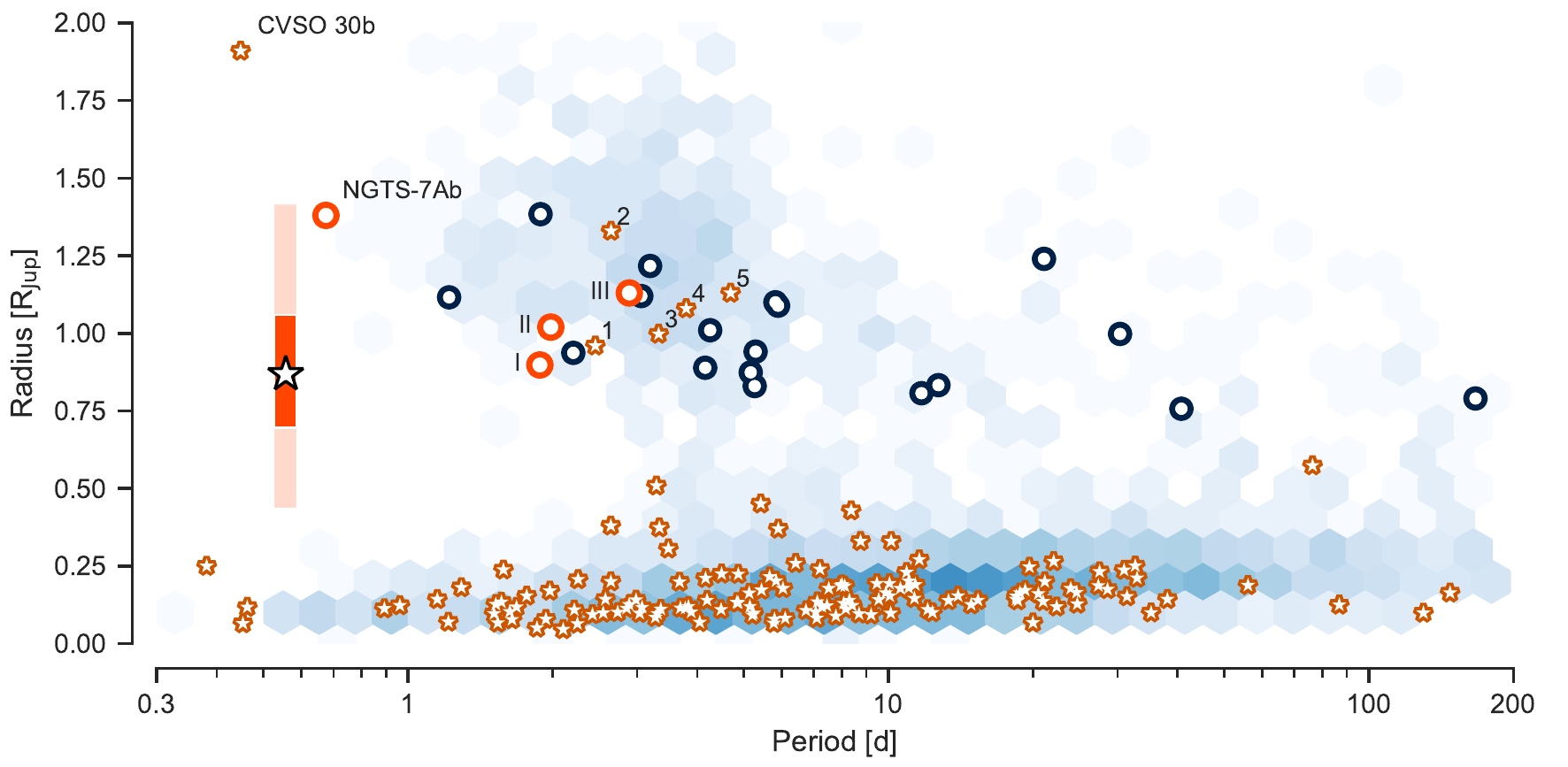}
	\caption{A period-radius diagram of the confirmed transiting extrasolar planets (\protect\url{exoplanet.eu}, \citealt{Schneider2011}, accessed 4.5.2019) and brown dwarfs \citep[collected from][]{Carmichael2019,Persson2019,Jackman2019} to date with periods between 0.3 and 200 days and radii between 0 and 2 \rjup . \tplanet is marked with a black-edged white star with 68\% and 99\% central posterior intervals for its true radius marked with dark and light orange shading, respectively. Planets around host stars with $T_\mathrm{Eff} < 4000K$ are marked with orange-edged stars, and the distribution of planets around hotter stars is shown as blue shading. Brown dwarfs around host stars with $T_\mathrm{Eff} < 4000K$ are marked with orange-edged circles, and the brown dwarfs around hotter hosts are marked with dark-blue-edged circles. The numbered planets and brown dwarfs are described in Sect.~\ref{sec:discussion}.}
	\label{fig:periodradius}
\end{figure*}

\tplanet offers a new puzzle for planetary system formation. If \tplanet is a planet, the lowest-mass possibility would make the object a hot Jupiter with an orbital period of only 0.56~days around a low mass star. Such planets, however, have never been observed before. In Fig.~\ref{fig:periodradius}, we show a period-radius diagram of the currently known transiting planets and brown dwarfs with periods between 0.3 and 200 days and radii between 0 and 2 \rjup, with the position of \tplanet marked with a star. \tplanet lies in a so-far unexplored parameter space. Further, if we remove all spectral hosts with $\teff > 4000$~K (non-M dwarfs), \tstar becomes unique and isolated. It is possible we are capturing the initial stages of a compact ultra-short period planet formation, before the planet can lose its atmosphere.

Given the uncertainty in the determination of the radius of \tplanet, and the degeneracy of the mass-radius relationship for planets and brown dwarfs more massive than 0.5~\mjup \citep[eg.][]{Baraffe2003}, there is a non-negligible probability that our target is high-mass planet or a brown dwarf with a mass in the interval $13\mjup < M < 75\mjup$. This would make \tstar an even more exciting system representing a unique and extreme case in the (disputed) "brown dwarf desert" \citep[][but see also \citealt{Carmichael2019} and \citealt{Persson2019}]{Marcy2000}, 
with the period of the transiting body falling between the 10.6~h orbital period of CVSO~30b \citep[][labelled in Fig.~\ref{fig:periodradius}]{VanEyken2012} and the 16.2~h orbital period of the recently discovered NGTS-7Ab \citep[][labelled in Fig.~\ref{fig:periodradius}]{Jackman2019}. Given the evidence that CSVO-30b is likely not a planet \citep{Yu2015,Lee2017}, \tplanet and NGTS-7Ab are the only currently known short-period massive objects of their kind orbiting M dwarfs.

In general, only a handful of massive transiting planets and brown dwarfs are known to orbit around M-dwarf host stars, and all except NGTS-7Ab have orbital periods greater than one day, as shown in Fig.~\ref{fig:periodradius}. The planets are 1) Kepler-45b \citep{Johnson2012}, 2) NGTS-1b \citep{Bayliss2018}, 3) HATS-6b \citep{Hartman2015}, 4) HATS-71Ab \citep{Bakos2018}, and 5) GJ~674b \citep{Bonfils2007}; and the brown dwarfs are I) LP~261-75b \citep{Society2017}, II) AD~3116b \citep{Gillen2017}, and III) NLTT~41135b \citep{Irwin2010} (excluding objects with periods larger than 200 days and radii larger than $2\,\rjup$, although the recent discovery of GJ~3512 \citep{Morales2019} with an orbital period of 204 days deserves to be mentioned). At the moment these objects or their host stars do not seem to share any clear common characteristics. NLTT~41135b and HATS-71b orbit a host in a binary system (HATS-71b likely), AD~3116 is a member of the Praesepe cluster, while HATS-6b would seem to be very normal warm Saturn orbiting a normal M1 star.

The existence of a massive planet or a brown dwarf at an orbital period of 0.56~d around an M3.5 dwarf star is very hard to explain using formation models based on core-accretion processes given the high migration rate of planet seeds around these low-mass stars \citep{Johansen2019}. 
These companions may be explained by disc fragmentation mechanisms at large separations from the parent star and later migration to close-in orbits \citep{Morales2019}. This model, however, is yet to be proven and \tplanet would become an excellent target for this endeavour before the massive planet or brown dwarf becomes engulfed by its parent star as predicted by the disc-fragmentation model \citep{Armitage2002}. 

\tstar is indeed a rare system among the thousands of transiting objects discovered to date because it offers an exclusive opportunity to constrain the formation models of planet and brown dwarf birth. The next step for this system is to determine the mass of \tplanet using radial velocity measurements. While \tstar is very faint, the measurements may be possible due to the low mass of the host star.
At the low-mass end ($M_\mathrm{p} \approx 0.5\,\mjup$), we can expect an RV semi-amplitude of 250 m/s,
while at the high-mass end ($M_\mathrm{p} \approx 75\,\mjup$) it could be around 35 km/s.

\begin{acknowledgements}
  First, we thank the anonymous referee for their helpful and constructive comments.
  We acknowledge financial support from the Agencia Estatal de Investigación of the Ministerio de Ciencia,
  Innovación y Universidades and the European FEDER/ERF funds through projects ESP2013-48391-C4-2-R, AYA2016-79425-C3-2-P, AYA2015-69350-C3-2-P
 This work is partly financed by the Spanish Ministry of Economics and Competitiveness through project ESP2016-80435-C2-2-R. NN acknowledges supports by JSPS KAKENHI Grant Numbers JP18H01265 and JP18H05439, and JST PRESTO Grant Number JPMJPR1775.
 JK acknowledges support by Deutsche Forschungsgemeinschaft (DFG) grants PA525/18-1 and PA525/19-1 within the DFG Schwerpunkt SPP 1992, Exploring the Diversity of Extra-solar Planets.
  This article is partly based on observations made with the MuSCAT2 instrument, developed by ABC, at Telescopio Carlos S\'anchez operated on the island of Tenerife by the IAC in the Spanish Observatorio del Teide. We acknowledge the use of public TESS Alert data from pipelines at the TESS Science Office and at the TESS Science Processing Operations Center. This work makes use of observations from the LCOGT network. Resources supporting this work were provided by the NASA High-End Computing (HEC) Program through the NASA Advanced Supercomputing (NAS) Division at Ames Research Center for the production of the SPOC data products. This work makes use of observations from the LCOGT network.
\end{acknowledgements}

\bibliographystyle{aa} 
\bibliography{toi_263}

\end{document}